\documentclass[hidelinks,prd,eqsecnum,amsmath,amssymb,nofootinbib,12pt,superscriptaddress]{revtex4-1}

\usepackage{dcolumn}
\usepackage{bm}
\usepackage{xcolor}
\usepackage[applemac]{inputenc}
\usepackage[spanish,english]{babel}
\usepackage{amsmath,amssymb,amsfonts,latexsym,cancel,amsthm,hyperref,bbm}
\usepackage{graphicx}
\usepackage{color}
\usepackage{soul}
\usepackage{ulem}
\usepackage{slashed}
\usepackage{braket}
\usepackage[mathscr]{euscript}
\usepackage{physics}
\usepackage{mathrsfs}
\usepackage{pgfplots}
\usepackage{float}
\usepackage{xfrac}
\usepackage{xcolor}
\usepackage{trimclip}
\usepackage{amsmath}
\usepackage{accents}
\definecolor{blue(pigment)}{rgb}{0.2, 0.2, 0.6}
\allowdisplaybreaks

\newcommand{\w}{\omega}

\newcommand{\be}{\begin{equation}}
\newcommand{\ee}{\end{equation}}
\newcommand{\bea}{\begin{eqnarray}}
\newcommand{\eea}{\end{eqnarray}}
\newcommand{\bes}
{\begin{subequations}}
\newcommand{\ees}{\end{subequations}}

\begin{document}
  \title{\bf Late-time behaviors of scalar field modes for a collapsing null shell spacetime and for the Unruh state in Schwarzschild spacetime}
 
    \author{Shohreh Gholizadeh Siahmazgi}\email{gholizs@wfu.edu}
    \affiliation{Department of Physics, Wake Forest University, Winston-Salem, NC, 27109, USA}

   \author{Paul R. Anderson}\email{anderson@wfu.edu}
    \affiliation{Department of Physics, Wake Forest University, Winston-Salem, NC, 27109, USA}

\author{Alessandro Fabbri}\email{afabbri@ific.uv.es}
    \affiliation{Departamento de F$\acute{\imath}$sica Te\'orica and IFIC, Universidad de Valencia-CSIC,\\
C. Dr. Moliner 50, 46100 Burjassot, Spain}
\begin{abstract}

   The behaviors of the modes for a massless minimally coupled scalar field are investigated for the Unruh state for Schwarzschild spacetime and the {\it in} vacuum state for a spacetime in which a spherically symmetric null shell collapses to form a nonrotating black hole.  In both cases there are two different sets of solutions to the mode equation that make up the state. For both spacetimes, one set of modes oscillates forever with no damping of the oscillations and the other set approaches zero at late times.  The difference between a mode that oscillates forever in the null-shell spacetime and the corresponding mode for the Unruh state vanishes as a power law in time. The modes that approach zero at late times also vanish at late times as a power law in time.  In all cases the power-law damping is preceded by a period of oscillations that appear to be due to quasi-normal modes.  
   
   \end{abstract}

    \date{\today}
    \maketitle

\newpage

\section{Introduction}

Hawking's original calculation predicting black hole evaporation~\cite{SHawking} was for black holes that form from collapse.  However, due to technical difficulties, most calculations of quantum effects in black hole spacetimes in four dimensions, 4D, have either made use of significant approximations or have involved eternal black holes which are mathematically much easier to work with.  For eternal black holes, it is the Unruh state~\cite{Unruh} that gives a flux of particles far from the black hole that is equivalent to that which occurs at late times for black holes that form from collapse.  

A question that is largely unexplored in four dimensions, 4D, is, what are the details of the late-time behaviors of the modes for both the Unruh state in an eternal black hole spacetime and the {\it in} state for a spacetime with a black hole that forms from collapse?  Here we address that question for the case of Schwarzschild spacetime and the case of a black hole that forms from the collapse of a spherically symmetric null shell.  A method to compute the stress-energy tensor for a massless minimally coupled scalar field in a collapsing null shell spacetime was given in~\cite{paper1}.  Part of the method involves the computation of the modes for both the {\it in} and Unruh states.

In this paper, we show the results of numerical computations of the spherically symmetric  modes for both the {\it in} state in the collapsing null shell spacetime and the Unruh state in Schwarzschild spacetime.  We restrict our attention to spherically symmetric modes since these are the easiest to work with in the collapsing null shell spacetime and they give the most important contribution to the Hawking radiation~\cite{Balbinot:2000iy}.  
We find that the {\it in} modes approach the Unruh modes at late times as a power law in time.  This is in contrast to the 2D case where the approach is exponential in time.  In the 4D, but not the 2D case, both sets of modes exhibit oscillatory behavior at intermediate times that is probably due to quasinormal modes.  

In Sec.~\ref{sec:background}, we review some basic properties of both Schwarzschild and the collapsing null shell spacetimes.  Our method to compute the {\it in} modes is summarized in Sec.~\ref{sec:method}.  Some results of our numerical computations are given in Sec.~\ref{sec:numerical}.  Section~\ref{sec:summary} contains a summary and our conclusions.  Throughout we use units such that $\hbar = c = G = 1$.

\section{Background}
\label{sec:background}

In the collapsing null shell spacetime, the metric inside the shell is the flat space metric
\be ds^2 = -dt^2 + dr^2 + r^2 d \Omega^2 \;, \label{metric-flat} \ee
and the metric outside the shell is the Schwarzschild metric
\be ds^2 = -\left( 1 - \frac{2 M}{r} \right) dt_s^2 + \left( 1 - \frac{2 M}{r} \right)^{-1}dr^2 + + r^2 d \Omega^2 \;. \label{metric-Sch} \ee
The radial coordinate $r$ is continuous across the shell.  The radial null coordinates inside the shell are
\bea v &=& t + r \;, \nonumber  \\
    u &=& t -r \;. \eea
Outside the shell they are
\bea v &=& t_s + r_* \;, \nonumber  \\
    u_s &=& t_s -r_* \;, \nonumber \\
    r_* &=& r + 2 M \log \left(\frac{r-2M}{2M} \right) \;.   \eea
The coordinate $v$ is continuous across the shell and the shell moves along the trajectory $v = v_0$.  On the trajectory~\cite{Sandro-book}
\be
u_s = u - 4M \log \left( \frac{v_H-u}{4 M} \right)  \;,
\label{us-u}
\ee
with $v_H \equiv v_0-4M$.
\begin{figure}[h]
\centering
\includegraphics[trim=6cm 20cm 5cm 0cm,clip=true,totalheight=0.2\textheight]{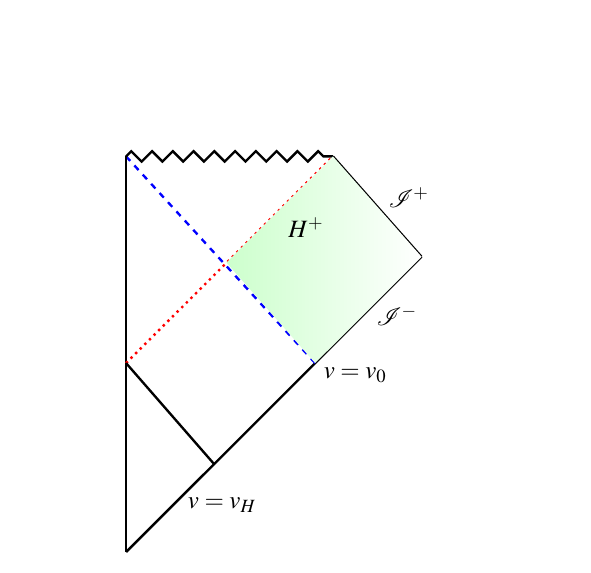}
\includegraphics[trim=0cm 0cm 0cm 0cm,clip=true,totalheight=0.16\textheight]{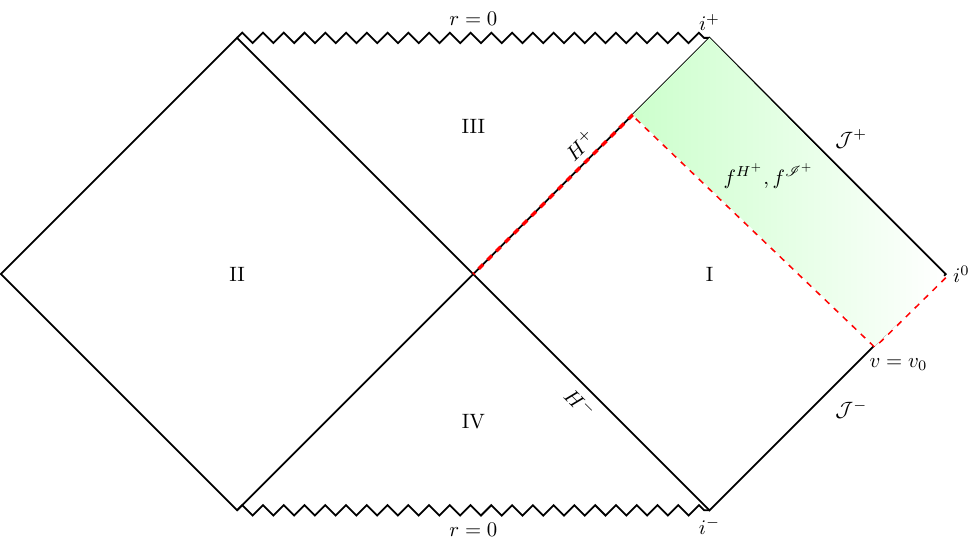}
\caption{The Penrose diagram on the left is for the null shell spacetime and that on the right is for Schwarzschild spacetime.  The shaded region in the diagram on the left is outside of the shell trajectory and outside of the event horizon.  The shaded region in the diagram on the right is the corresponding region in Schwarzschild spacetime.  The red dotted line in the right-hand diagram shows the Cauchy surface that is used to compute the matching coefficients. }
\label{fig:I1}
\end{figure}

\section{Computation of the modes}
\label{sec:method}

The wave equation for the massless minimally coupled scalar field is $\Box \phi = 0$ and this is also the equation for the modes of the quantum field. 
All of the mode functions that we work with can be written in the form
\be f_{\w \ell m} = \frac{Y_{\ell m}(
\theta, \phi)}{r \sqrt{4 \pi \w}}  \psi_{\w \ell}(t,r) \;, \label{f-general} \ee
where the functions $Y_{\ell, m}$ are the usual spherical harmonics and $\w \ge 0$.

The {\it in} vacuum state is defined such that everywhere on past null infinity
\be f^{\rm in}_{\w \ell m} \sim  e^{-i \w v}  \; .  \label{in-state-def} \ee
It is further required that inside the shell the modes be regular at $r = 0$.
The result is that inside the shell~\cite{paper1}
\be \psi^{\rm in}_{\w \ell} =  C_{\ell} \, e^{-i \w t}\, \w r  j_\ell(\w r) \;, \label{psi-in-a} \ee
where $C_{\ell}$ is a normalization constant and $j_\ell $ is a spherical Bessel function.  
We use the usual scalar product denoted by 
$(\phi_1,\phi_2)$ to normalize the modes by requiring that 
\be (f^{\rm in}_{\w \ell m}(x), f^{\rm in}_{\w' \ell' m'}(x)) = \delta_{\ell,\ell'}  \delta_{m, m'} \delta(\w-\w')\;. \label{normalization-condition}\ee
The scalar product is independent of the Cauchy surface used.  For the null-shell spacetime it can be seen from Fig.~\ref{fig:I1} that past null infinity is a Cauchy surface.  
For $\ell = 0$, it is straightforward to show that $C_{0} = -2 i $ and
\be \psi^{in}_{\w 0} = e^{-i \w v} - e^{-i \w u} \;. \label{psi-in-0} \ee

Outside the shell the {\it in} modes have a much more complicated form.  To compute them we first expand them in terms of a complete set of modes in Schwarzschild spacetime that can be obtained using separation of variables.  The complete set we choose consists of the union of two sets of modes that we call $f^{H^+}$ and $f^{\mathscr{I}^+}$ which are defined by their behaviors on the future horizon and future null infinity respectively.  On the future horizon 
\be \psi^{H^+}_{\w \ell} \to e^{-i \w v} \;, \ee
and on future null infinity
\be \psi^{\mathscr{I}^+}_{\w \ell} \to e^{-i \w u_s}  \;. \ee

These modes are orthonormal in terms of the usual scalar product in Schwarzschild spacetime.  This orthonormality can be used to compute the matching coefficients. 
Since the matching is done in pure Schwarzschild spacetime, we use a Cauchy surface for the region outside of the past and future horizons.  It is shown in Fig.~\ref{fig:I1} and consists of the part of the future horizon from $v = -\infty$ to $v = v_0$, the trajectory of the null shell $v = v_0$ in the region outside the future horizon, and the part of past null infinity that goes from $v = v_0$ to $v = \infty$.  The part on the future horizon does not causally affect the shaded region which is outside the null shell trajectory and outside the future horizon, so any values for $\psi^{\rm in}_{\w \ell}$ on that part of the Cauchy surface can be chosen.

The general form of the expansion is
\bea
f^{in}_{\w \ell m} &=& \sum_{\ell'} \sum_{m'} \int_{0}^{\infty} d \w'
\Big[ A^{\mathscr{I^+}}_{\w \ell m \w' \ell' m'} f^{\mathscr{I}^+}_{\w' \ell' m'} + B^{\mathscr{I^+}}_{\w \ell m \w' \ell' m'} (f^{\mathscr{I}^+}_{\w' \ell', m'})^{*} \nonumber \\
& &  + A^{H^+}_{\w \ell m \w' \ell' m'} f^{H^+}_{\w' \ell' m'} + B^{H^+}_{\w \ell m \w'\ell' m'} (f^{H^+}_{\w' \ell' m'})^{*}\Big]\;. \label{General-in-modes-1}
\eea
In~\cite{paper1} it was shown that this can be simplified significantly with the result
\be
f^{in}_{\w \ell m} =  \frac{Y_{\ell m}}{r \sqrt{4 \pi}} \int_{0}^{\infty} \frac{d \w'}{\sqrt{\w'}}
\Big[ A^{\mathscr{I^+}}_{\w \w' \ell} \psi^{\mathscr{I}^+}_{\w'\ell} + B^{\mathscr{I^+}}_{\w \w' \ell} (\psi^{\mathscr{I}^+}_{\w' \ell})^{*}
 + A^{H^+}_{\w \w' \ell} \psi^{H^+}_{\w' \ell} + B^{H^+}_{\w \w'\ell} (\psi^{H^+}_{\w' \ell})^{*}\Big]\;. \label{General-in-modes-2}
\ee
We drop the subscripts for $\ell$ and $m$ in what follows since we are only considering spherically symmetric modes which have $\ell = m = 0$.  
We separate the contributions of $\psi^{\rm in}_\w$ to the scalar products for the matching coefficients into two parts that we call $(\psi^{\rm in}_\w)_v$ and $(\psi^{\rm in}_\w)_u$.  On the part of the Cauchy surface that is on $H^{+}$, we choose $(\psi^{\rm in}_\w)_v = e^{-i \w v_0}$  and $(\psi^{\rm in}_\w)_u = - e^{-i \w v_H}$.  On the part containing the null shell,  $(\psi^{\rm in}_\w)_v = e^{-i \w v_0}$  and $(\psi^{\rm in}_\w)_u = - e^{-i \w u(u_s)}$.  On the part that is on past null infinity, $(\psi^{\rm in}_\w)_v = e^{-i \w v}$,
Since $u = -\infty$ on this surface,  $(\psi^{\rm in}_\w)_u$ oscillates an infinite number of times and is thus effectively zero.  The result is 
\bes \bea
\big(A^{H^+}_{\w \w'}\big)_v&=&\frac{i}{2 \pi} \sqrt{\frac{\w'}{\w}} \frac{1}{F_L^{*}(\w')}\frac{e^{i(\w'-\w)v_0}}{\w'-\w+i \epsilon},\\
\big(B^{H^+}_{\w \w'}\big)_v&=& - \frac{i}{2 \pi} \sqrt{\frac{\w'}{\w}} \frac{1}{F_L(\w')}\frac{e^{-i(\w+\w')v_0}}{\w'+\w-i \epsilon},\label{ABH-v-part} \\
\big(A^{\mathscr{I}^+}_{\w \w'}\big)_v&=& - \frac{i}{2 \pi} \sqrt{\frac{\w'}{\w}} \frac{F_R^{*}(\w')}{F_L^{*}(\w')}\frac{e^{-i(\w-\w')v_0}}{\w'-\w+i \epsilon}, \\
\big(B^{\mathscr{I}^+}_{\w \w'}\big)_v&=& \frac{i}{2 \pi} \sqrt{\frac{\w'}{\w}} \frac{F_R(\w')}{F_L(\w')}\frac{e^{-i(\w+\w')v_0}}{\w'+\w-i \epsilon}.\label{ABI-v-part}
\eea \label{ch5-ABHIv}\ees
where the scattering parameters $F_R$ and $F_L$ are defined in Eq.~\eqref{near-horizon}.
We also find that 
\bes \bea
\big(A^{H^+}_{\w \w'}\big)_u&=& - \frac{1}{2 \pi} \sqrt{\frac{\w }{\w'}}  \int_{-\infty}^{v_H} du \,  e^{-i \w u}  \psi^{H^{+} *}_{\w' 0}(u_s(u),v_0),\label{ch5-AHu}\\
\big(B^{H^+}_{\w \w'}\big)_u&=& \frac{1}{2 \pi} \sqrt{\frac{\w}{ \w'}}  \int_{-\infty}^{v_H} du \,e^{-i \w u} \psi^{H^{+}}_{\w' 0} (u_s(u),v_0),\label{ch5-BHu}\\
\big(A^{\mathscr{I}^+}_{\w \w'}\big)_u&=&\frac{i}{2 \pi} \frac{1}{\sqrt{\w \w'}} \int_{-\infty}^{v_H} du \, e^{-i \w u} \partial_u \psi^{\mathscr{I}^+  *}_{\w' 0}(u_s(u),v_0),\label{ch5-AIu}\\
\big(B^{\mathscr{I}^+}_{\w \w'}\big)_u&=&- \frac{i}{2 \pi} \frac{1}{\sqrt{\w \w'}} \int_{-\infty}^{v_H} du \,e^{-i \w u} \partial_u \psi^{\mathscr{I}^+}_{\w' 0}(u_s(u),v_0).\label{ch5-BIu}
\eea \label{ch5-ABHIu}\ees
The results in~\eqref{ch5-ABHIv} were derived in~\cite{paper1}.\footnote{The terms in $A^{H^+}_{\w \w'}$ and $B^{H^+}_{\w \w'}$ that come from the evaluation of the scalar product on the future horizon and therefore don't contribute to solutions in the region of interest have been omitted.}  The results in~\eqref{ch5-ABHIu} were obtained using a straightforward extension of that derivation.
Note that the integrals in~\eqref{ch5-ABHIu} are to be computed on the null-shell trajectory and since $\psi^{H^{+} *}_{\w'}$ and $\psi^{\mathscr{I}^{+} *}_{\w' 0}$ do not have analytic forms on the null-shell trajectory, these integrals have to be evaluated numerically. 

The modes for the Unruh state in Schwarzschild spacetime consist of the modes $f^{\mathscr{I}^-}$ for which on past null infinity
\be \psi^{\mathscr{I}^-}_{\w} \to e^{-i \w v} \;, \ee
and the Kruskal modes $f^K$ for which on the past horizon
\be \psi^K_{\w} \to e^{-i \w U} \;, \ee
with
\be U = - \frac{e^{-\kappa u_s}}{\kappa} \;.  \label{U-Kruskal}  \ee
The Kruskal modes in turn can be expanded in terms of the modes $f^{H^-}$ for which on the past horizon  
\be \psi^{H^-}_{\w} \to e^{-i \w u_s}  \;. \ee
The expansion takes the form
\be f^K_{\w} =  \int_0^\infty d \w' \, \left[  \alpha^K_{\w \w'} f^{H^{-}}_{\w'} + \beta^K_{\w \w'}   f^{H^{-}*}_{\w'} \right]  \;, \label{fK} \ee
with
\bes \bea \alpha^K_{\w_K \w'} &=&  \frac{1}{2 \pi} \sqrt{\frac{\w'}{\w_K}} (4 M)^{1+i 4 M \w'}  \int_{-\infty}^0 d U_K e^{-i \w_K U_K} (-U_K)^{-1 - i 4 M \w'} \nonumber \\
&=& \frac{1}{2 \pi} \sqrt{\frac{\w'}{\w_K}} \, (4M)^{1 +i 4 M \w'} \frac{\Gamma(\delta - i 4 M \w)}{ (-i \w_K + \epsilon)^{-i 4 M \w'}} \;, \label{alphaK} \\
\beta^K_{\w_K \w'} &=&  \frac{1}{2 \pi} \sqrt{\frac{\w'}{\w_K}} (4 M)^{1-i 4 M \w'}  \int_{-\infty}^0 d U_K e^{-i \w_K U_K} (-U_K)^{-1 + i 4 M \w'} \nonumber \\
&=& \frac{1}{2 \pi} \sqrt{\frac{\w'}{\w_K}} \, (4M)^{1 -i 4 M \w'} \frac{\Gamma(\delta + i 4 M \w)}{ (-i \w_K + \epsilon)^{i 4 M \w'}} \;. \label{betaK}
\eea \label{alphaK-betaK} \ees

The solutions for the modes $\psi_w^{(\mathscr{I}^+, H^+)}$ can be obtained using separation of variables and writing
\be \psi_w^{(\mathscr{I}^+, H^+)}(t,r) = e^{-i \w t} \chi_w^{(\mathscr{I}^+, H^+)}(r) \;, \ee
with
\bes \bea & & \frac{d^2 \chi}{d r^2_*} + (\w^2 - V_{\rm eff}) \chi = 0 \;, \\
   & &  V_{\rm eff} = \left(1-\frac{2 M}{r} \right) \left( \frac{2M}{r^3} + \frac{\ell (\ell+1)}{r^2} \right) \;, \label{Veff-def} \eea \label{radial-mode-eq}\ees
  where the dependence on $\ell$ has been shown for completeness.
  
   For the numerical computations, we find two linearly independent solutions to~\eqref{radial-mode-eq} which we denote by $\chi^\infty_R$ and $\chi^\infty_L$.  They are defined by their behaviors at large values of $r$ which are 
\be \chi^\infty_R \to e^{i \w r_*} \;, \qquad \chi^\infty_L \to e^{-i \w r_*} \;. \label{large-distance} \ee
Since the effective potential in the radial mode equation vanishes at the horizon, we have the general forms
\bea \chi^\infty_R & \to & E_R e^{i \w r_*} + F_R e^{-i \w r_*} \;, \nonumber \\
     \chi^\infty_L & \to & E_L e^{i \w r_*} + F_L e^{-i \w r_*} \;. \label{near-horizon} \eea
The scattering coefficients $E_R, \; F_R, \; E_L$, and $\; F_L$ are computed numerically.

We then find that~\cite{paper1}
\bea
\psi^{H^+}_{\w}=\frac{\chi_L^{\infty}(r)}{F_L(\w)}e^{-i\w t_s}\label{psiHplus}
\eea
and \bea\psi^{\mathscr{I}^+}_{\w}=\Big(\chi_R^{\infty}(r)-\frac{F_R(\w)}{F_L(\w)}\chi^{\infty}_{L}\Big)e^{-i\w t_s}.\label{psiIplus}
\eea

As mentioned above, the integrals in~\eqref{ch5-ABHIu} have to be computed numerically. The first step is to convert them to integrals over $r$ noting that $v = v_0$ for the surface being integrated over so that $u = v_0 - 2r$.  Then the integration variable is switched to $s = M^{-1}(r-2M)$.  

Examination of~\eqref{Veff-def} shows that $V_{\rm eff}$ vanishes in the limits $s \to 0$ and $s \to \infty$.  This means that for any value of $\w$ it is possible close to the horizon to find some value of $s_1 >0$ such that for $0 \le s \le s_1$, $\w^2 \gg |V_{\rm eff}(s)|$. Similarly it is possible far from the horizon to find some $s_2 > s_1$ such that $\w^2 \gg |V_{\rm eff}(s)|$ for $s_2 \le s < \infty$.  In both cases it is possible to find analytic expressions for the integrands in~\eqref{ch5-ABHIu}.
In the near-horizon limit, $\chi_R^{\infty}(\w,s)$ and $\chi_L^{\infty}(\w,s)$ are given by ~\eqref{near-horizon} and in the large-distance limit,
$\chi_R^{\infty}(\w,s)$ and $\chi_L^{\infty}(\w,s)$ are given by~\eqref{large-distance}. For intermediate values of $s$, numerical computations must be performed. 

Thus the integrals in ~\eqref{ch5-ABHIu} can be broken into the three parts 
\be \int_0^{s_1} ds +  \int_{s_1}^{s_2} ds +  \int_{s_2}^\infty ds \;, \ee
The contributions of the first integral to the matching coefficients can be written in terms of lower incomplete gamma functions with the result
\bes \bea
(A^{H^+}_{\w\w'})_{u \,1}&=&-\frac{e^{i\gamma_A}}{2i\w\pi}\sqrt{\frac{\w}{\w'}}e^{i\w' (2-2\log 2)}\Big(e^{2i\w s_1}-1\Big)\nonumber \\ &&
-\frac{e^{i\gamma_A}}{\pi}\sqrt{\frac{\w}{\w'}}\frac{F_R}{E_R}e^{-i\w' (2-2\log 2)}\Big(\frac{-1}{2i(\w-\w')}\Big)^{-4i\w'+1}\gamma\Big(-4i\w'+1,t_1\Big)\label{ch5-AHsamll} \\
(B^{H^+}_{\w\w'})_{u \,1}&=&\frac{e^{i\gamma_B}}{2i\w\pi}\sqrt{\frac{\w}{\w'}}e^{-i\w' (2-2\log 2)}\Big(e^{2i\w s_1}-1\Big)\nonumber \\ &&
+\frac{e^{i\gamma_B}}{\pi}\sqrt{\frac{\w}{\w'}}\frac{F_R^*}{E_R^*}e^{i\w' (2-2\log 2)}\Big(\frac{-1}{2i(\w+\w')}\Big)^{4i\w'+1}\gamma\Big(4i\w'+1,t_2\Big) \;, \\
(A^{\mathscr{I}^+}_{\w\w'})_{u \,1}&=&
-\frac{\w'e^{-i\w'(2-\log 2)}}{\pi\sqrt{\w\w'}}\frac{e^{i\gamma_A}}{(2i(\w'-\w)+\epsilon)^{-4i\w'}}\gamma(-4i\w'+\delta,t_1)\nonumber \\ &&
-\frac{\w'e^{-i\w'(2-\log 2)}}{\pi\sqrt{\w\w'}}\frac{e^{i\gamma_A}}{(2i(\w'-\w)+\epsilon)^{-4i\w'+1}}\gamma(-4i\w'+1+\delta,t_1),\\
(B^{\mathscr{I}^+}_{\w\w'})_{u \,1}&=&
+\frac{\w'e^{i\w'(2-\log 2)}}{\pi\sqrt{\w\w'}}\frac{e^{i\gamma_B}}{(-2i(\w'+\w)+\epsilon)^{4i\w'}}\gamma(4i\w'+\delta,t_2)\nonumber \\ &&
+\frac{\w'e^{i\w'(2-\log 2)}}{\pi\sqrt{\w\w'}}\frac{e^{i\gamma_B}}{(-2i(\w'+\w)+\epsilon)^{4i\w'+1}}\gamma(4i\w'+1+\delta,t_2) \;,
\eea \ees
where $t_1 \equiv -2i(\w-\w')s_1$ and $t_2 \equiv -2i(\w+\w')s_1$.
The contributions of the third integral to the matching coefficients can be written in terms of upper incomplete gamma functions with the result
\bes \bea  
\big(A^{H^+}_{\w \w'}\big)_{u \,3}&=& \frac{e^{4i\w'}}{\pi}\sqrt{\frac{\w}{\w'}}\frac{1}{F_L^*(\w')}\frac{e^{2i\w s_2}}{2i\w-\epsilon},\\
\big(B^{H^+}_{\w \w'}\big)_{u \,3}&=& \frac{e^{-4i\w'}}{\pi}\sqrt{\frac{\w}{\w'}}\frac{1}{F_L(\w')}\frac{e^{2i\w s_2}}{2i\w-\epsilon} \;, \\
(A_{\w\w'0}^{\mathscr{I}^+})_{u \,3}&=&
-\frac{\w'e^{-i\w'(2-\log 2)}}{\pi\sqrt{\w\w'}}\frac{e^{i\gamma_A}}{(2i(\w'-\w)+\epsilon)^{-4i\w'}}\Gamma(-4i\w',t_3)\nonumber \\ &&
-\frac{\w'e^{-i\w'(2-\log 2)}}{\pi\sqrt{\w\w'}}\frac{e^{i\gamma_A}}{(2i(\w'-\w)+\epsilon)^{-4i\w'+1}}\Gamma(-4i\w'+1,t_3) \;, \\
(B_{\w\w'0}^{\mathscr{I}^+})_{u \,3}&=&
\frac{\w'e^{i\w'(2-\log 2)}}{\pi\sqrt{\w\w'}}\frac{e^{i\gamma_B}}{(-2i(\w'+\w)+\epsilon)^{4i\w'}}\Gamma(4i\w',t_4)\nonumber \\ &&
+\frac{\w'e^{i\w'(2-\log 2)}}{\pi\sqrt{\w\w'}}\frac{e^{i\gamma_B}}{(-2i(\w'+\w)+\epsilon)^{4i\w'+1}}\Gamma(4i\w'+1,t_4),
\eea \ees
where $t_3 \equiv 2i(\w'-\w)s_2$ and $t_4 \equiv -2i(\w'+\w)s_2+\epsilon s_2$.

\section{Numerical results}
\label{sec:numerical}

To find $f^{\text{in}}_v$, one can use the expansion in \eqref{General-in-modes-2} with  the matching coefficients in \eqref{ch5-ABHIv}. Using ~\eqref{psiHplus} and \eqref{psiIplus}, together with the identity  $|F_L|^2-|F_R|^2=1$\footnote{This identity can be derived using the properties of the scattering coefficients given in \cite{rigorous}.}, we find
\bea
f^{\text{in}}_{v}&=&\frac{Y_{00}}{r\sqrt{4\pi \w}}\Bigg\{\chi_L^{\infty}e^{-i\w t_s}
-\frac{i}{2\pi}\int_{0}^{\infty}\frac{F_R^*(\w')}{F_L^*(\w')}\frac{e^{-i(\w-\w')v_0}}{\w'-\w+i\epsilon}\chi_R^{\infty}e^{-i\w' t_s}d\w'\nonumber \\ &&
+\frac{i}{2\pi}\int_0^{\infty}\frac{F_R(\w')}{F_L(\w')}\frac{e^{-i(\w'+\w)v_0}}{\w'+\w-i\epsilon}\chi_L^{\infty}e^{i\w' t_s}d\w'\Bigg\}.\label{IVfinv}
\eea
To avoid dealing with the singularity at $\w'=\w$ in the first integral, one can subtract off a term in which everything but the singular factor is evaluated $\w' = w$.  A similar term can be subtracted from the second integral.  When the two subtraction terms are added back, they can be computed analytically.  The result is
\bes \bea
f^{\text{in}}_{v}&=&\frac{Y_{00}}{r\sqrt{4\pi \w}}\Big\{\chi_L^{\infty}-\frac{F^*_R(\w)}{F^*_L(\w)}\chi_R^{\infty}(\w,r)\Big\}e^{-i\w t_s}+I_1+I_2,\label{IVfinv2} \;, \\
I_1&=& \frac{Y_{00}}{r\sqrt{4\pi\w}}  \Bigg\{
-\frac{i}{2\pi}\int_{0}^{\infty}\Big(\frac{F^*_R(\w')}{F^*_L(\w')}\frac{e^{-i(\w-\w')v_0}}{\w'-\w+i\epsilon}\chi_R^{\infty}(\w',r)e^{-i\w' t_s}\nonumber \\ &&
-\frac{F^*_R(\w)}{F^*_L(\w)}\frac{\chi_R^{\infty}(\w,r)}{\w'-\w+i\epsilon}e^{-i\w' t_s}
\Big)d\w'
\Bigg\} \;, \label{I1} \\
I_2&=&\frac{Y_{00}}{r\sqrt{4\pi\w}}  \Bigg\{\frac{i}{2\pi}\int_{0}^{\infty}\Big(\frac{F_R(\w')}{F_L(\w')}\frac{e^{-i(\w+\w')v_0}}{\w'+\w-i\epsilon}\chi_L^{\infty}(\w',r)e^{i\w' t_s}d\w'\nonumber \\ &&
-\frac{F_R^*(\w)}{F_L^*(\w)}\frac{\chi_R^{\infty}(\w,r)}{\w'+\w-i\epsilon}e^{i\w' t_s}
\Big)d\w'\Bigg\}.\label{I2}
\eea \ees 

Comparison with~\eqref{psiIplus} shows that the first term in~\eqref{IVfinv} is equal to $f^{\mathscr{I}^-}$. As mentioned above, the set of $f^{\mathscr{I}^-}$ modes is one of the two sets that constitute the Unruh state.
Thus, the sum $I_1+I_2$ gives the difference between the {\it in} state mode $f^{\text{in}}_{v}$ and the corresponding Unruh state mode $f^{\mathscr{I}^-}$.  We have computed $I_1+I_2$ numerically for several values of $\w$ at the fixed spatial point $r = 3M$ for a large range of times $t_s$. 

Numerical results for the frequency $M\w=0.5$ are shown in Fig.~\ref{fig:I2}.  The linear scale plots clearly show that the difference vanishes at late times.  The log-log plots show a very interesting structure.  The same structure occurs for the $f^{\rm in}_u$ modes and the $f^K$ modes.  It consists of contributions that appear to be quasinormal modes followed by a power-law tail which falls off in a way that is consistent with $t_s^{-3}$.  More details are given below.  
\begin{figure}[h]
\centering
\includegraphics[trim=0cm 0cm 0cm 0cm,clip=true,totalheight=0.22\textheight]{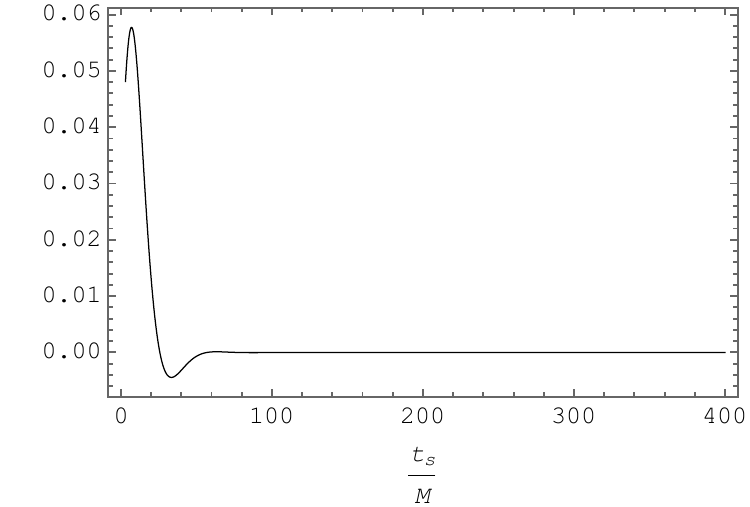}
\includegraphics[trim=0cm 0cm 0cm 0cm,clip=true,totalheight=0.22\textheight]{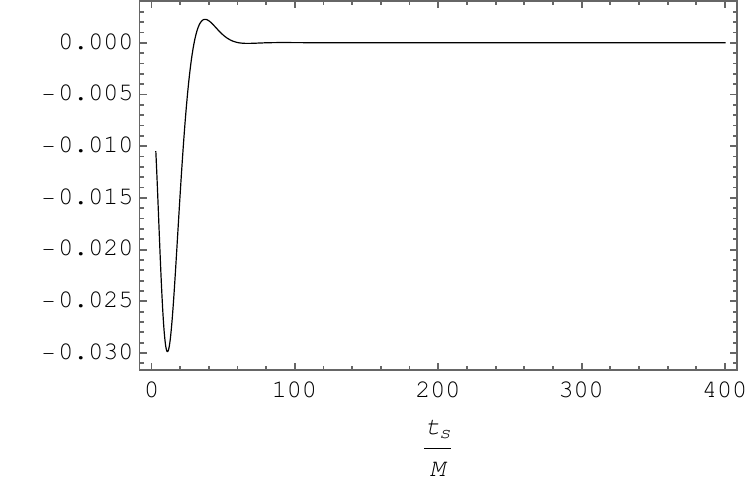}
\includegraphics[trim=0cm 0cm 0cm 0cm,clip=true,totalheight=0.22\textheight]{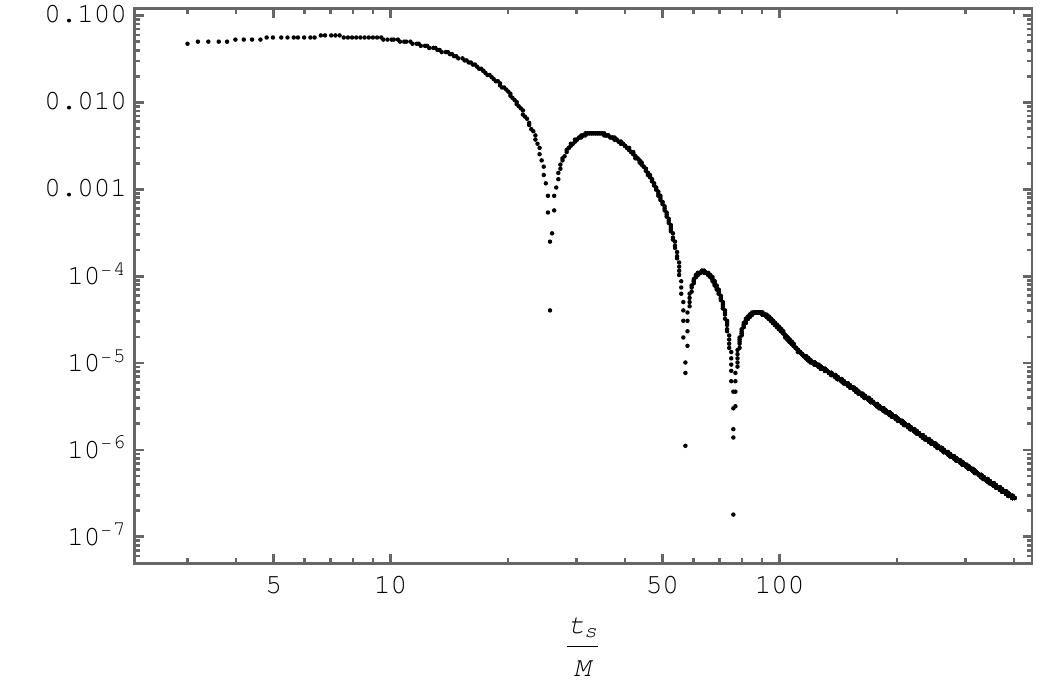}
\includegraphics[trim=0cm 0cm 0cm 0cm,clip=true,totalheight=0.22\textheight]{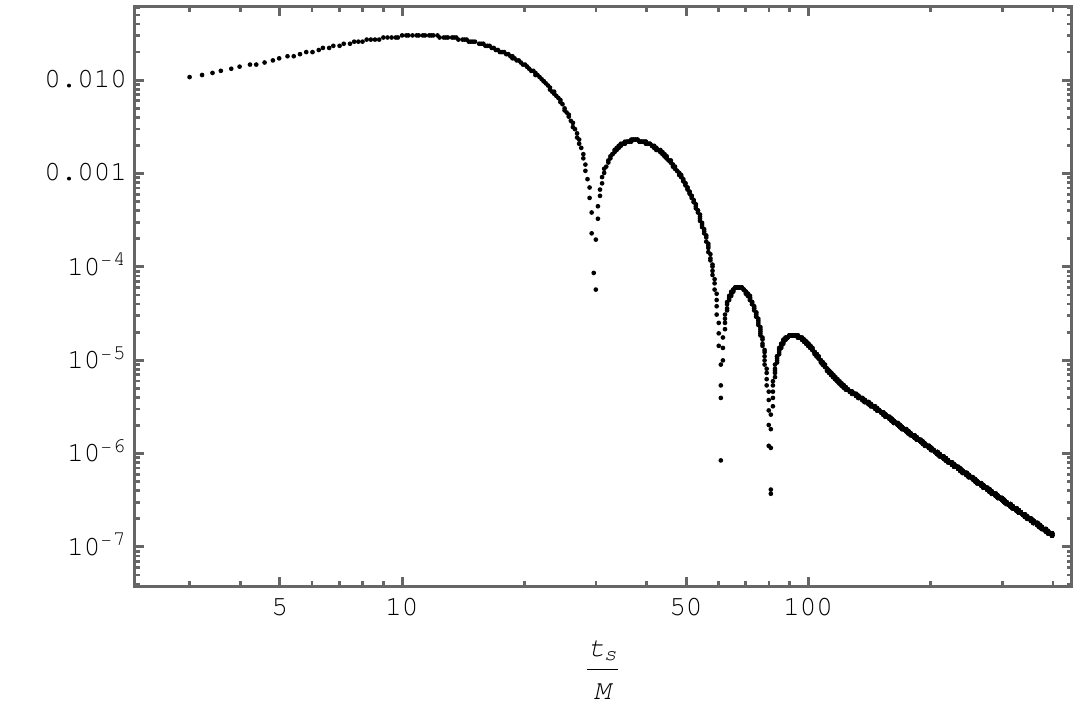}
\caption{The real (top-left) and imaginary (top-right) parts of $[(\psi^{\text{in}}_\w)_v-\psi^{\mathscr{I}^-}_\w]$ for the frequency $M\w=0.5$ and space point $r=3M$ are plotted as a function of the time $t_s$. At late times, $(\psi^{\text{in}}_\w)_v$ approaches $\psi^{\mathscr{I}^-}_\w$. The log-log plots for the real (bottom-right) and imaginary (bottom-left) parts of $[(\psi^{\text{in}}_\w)_v-\psi^{\mathscr{I}^-}_\w]$ show that this approach is a power law.}
\label{fig:I2}
\end{figure}

It is worth noting that all of this complex behavior is due to the fact that the mode functions are scattered due to the existence of an effective potential in the radial mode equation~\eqref{radial-mode-eq}.  If this potential is zero as it is in two dimensions, then~\cite{paper1} one finds in the region outside the shell and the horizon that
\be (\psi^{\rm in}_\w)_v = \psi^{\mathscr{I}^-}_\w = e^{-i \w v} \;. \ee

We next investigate the second set of modes that constitute the {\it in} state, $f^{\text{in}}_u$, and the Kruskal modes $f^{K}$ that make up part of the Unruh state. 
To find $f^{in}_u$, one needs to use the method discussed in the previous section to numerically compute the matching coefficients $(A^{H^+}_{\w \w'})_u$, $(B^{H^+}_{\w\w'})_u$, $(A^{\mathscr{I}^+}_{\w\w'})_u$, and $(B^{\mathscr{I}^+}_{\w\w'})_u$.  Then  the result is used to find $f^{\text{in}}_u$ using ~\eqref{General-in-modes-2}. Equations~\eqref{U-Kruskal} and~\eqref{alphaK-betaK}, are used to compute $f^K$. In both cases, the numerical computation has been done for a fixed frequency $\w=0.5$ and a fixed spatial coordinate $r=3M$. The results are depicted in Fig. 3. It can be seen that at sufficiently late times, the $f^{\rm in}_u$ and $f^K$ modes both approach zero.  For the  $f^K$ mode and the real part of the $f^{\text{in}}_u$ mode, this is as a power law consistent with $t_s^{-3}$. This is also very likely the late-time behavior of the imaginary part of  the $f^{\text{in}}_u$ mode.  However our numerical calculations are not accurate enough at late times to show this.

 The patterns in the log-log plots of Figures~2 and 3 are very similar.  At intermediate times there are a few damped oscillations followed by a power law tail. 
 This behavior is very similar to the late-time behavior found for the classical massless minimally-coupled scalar field in ~\cite{gcpp, burko-ori, barack}. The late-time power-law damping in that case also goes like $t_s^{-3}$ for spherically symmetric perturbations.
 
 The damped oscillations for the classical field are attributed to quasi-normal modes with complex frequencies.  These modes are exponentially damped in time.  This type of effect is also seen in the behavior of gravitational radiation emitted by a black hole after the ``merger" phase and during the ``ring-down phase''. A study of quasinormal modes and the power-law tail was first done in ~\cite{Leaver}.

Theoretical evidence for the presence of quasinormal modes interacting with Hawking radiation in an analog black hole model was found in~\cite{Jacquet} by studying the results of numerical quantum mechanics calculations of the density-density correlation function for the system using the truncated Wigner method.  However, to our knowledge the oscillations due to quasi-normal modes and the late-time power law tail that we find via direct calculations of the modes of a quantum field have not been seen in previous quantum field theory in curved space calculations.  The calculations for a classical massless minimally coupled scalar field in black hole spacetimes mentioned above involve the use of nontrivial initial data that is then evolved numerically.  This is also effectively what occurs for the $f^{\rm in}_v$, $f^{\rm in}_u$, and $f^K$ modes.  To see this note that these modes are all written in terms of packets of modes that are positive frequency with respect to $t_s$ and the complex conjugates of these modes.  The coefficients in these packets are determined by the behaviors of the $f^{\rm in}_v$, $f^{\rm in}_u$, and $f^K$ modes on the Cauchy surfaces where these coefficients are computed.  These behaviors  are mathematically the same as initial data for non-compact perturbations. While most or all of the classical calculations were carried out with compact initial data, it was shown in~\cite{AAG} that the same type of behavior also occurs classically for non-compact initial data. 

It is again useful to compare with the 2D case where the effective potential is zero.  Then in the region outside the shell and the event horizon one finds for the spherically symmetric modes that
\bea (\psi^{\rm in}_\w)_u &=& e^{-i \w u(u_s)}  \;, \nonumber \\
     \psi^K_\w &=& e^{-i \w U(u_s)} \;, \eea 
where $u(u_s) $ is given by the inverse of~\eqref{us-u} which is~\cite{GoodAndersonEvans}
\be u = v_H - \frac{1}{\kappa} W\left[ e^{(\kappa(v_H-u_s)} \right] \;,  \ee
with $W$ the product log function and $U(u_s)$ given by~\eqref{U-Kruskal}.
Since the shell trajectory $v = v_0$ and hence $v_H = v_0 - 4 M$ are arbitrary, we can without loss of generality set $v_H = 0$.  Then, since $W(x) = x - x^2 + O(x^3)$, it is easy to see that $u(u_s)$ approaches $U(u_s)$ exponentially in terms of $u_s$.  Further, since $U \to 0$ in the limit $u_s \to \infty$, both $(\psi^{\rm in}_\w)_u \to \psi^K_\w \to 1$ in this limit.  Thus scattering effects both slow down the approach of $(\psi^{\rm in}_\w)_u$ to $\psi^K_\w$  at late times and cause $(\psi^K_\w)_u$ to approach zero rather than $1$.
\begin{figure}[h]
\centering
\includegraphics[trim=0cm 0cm 0cm 0cm,clip=true,totalheight=0.22\textheight]{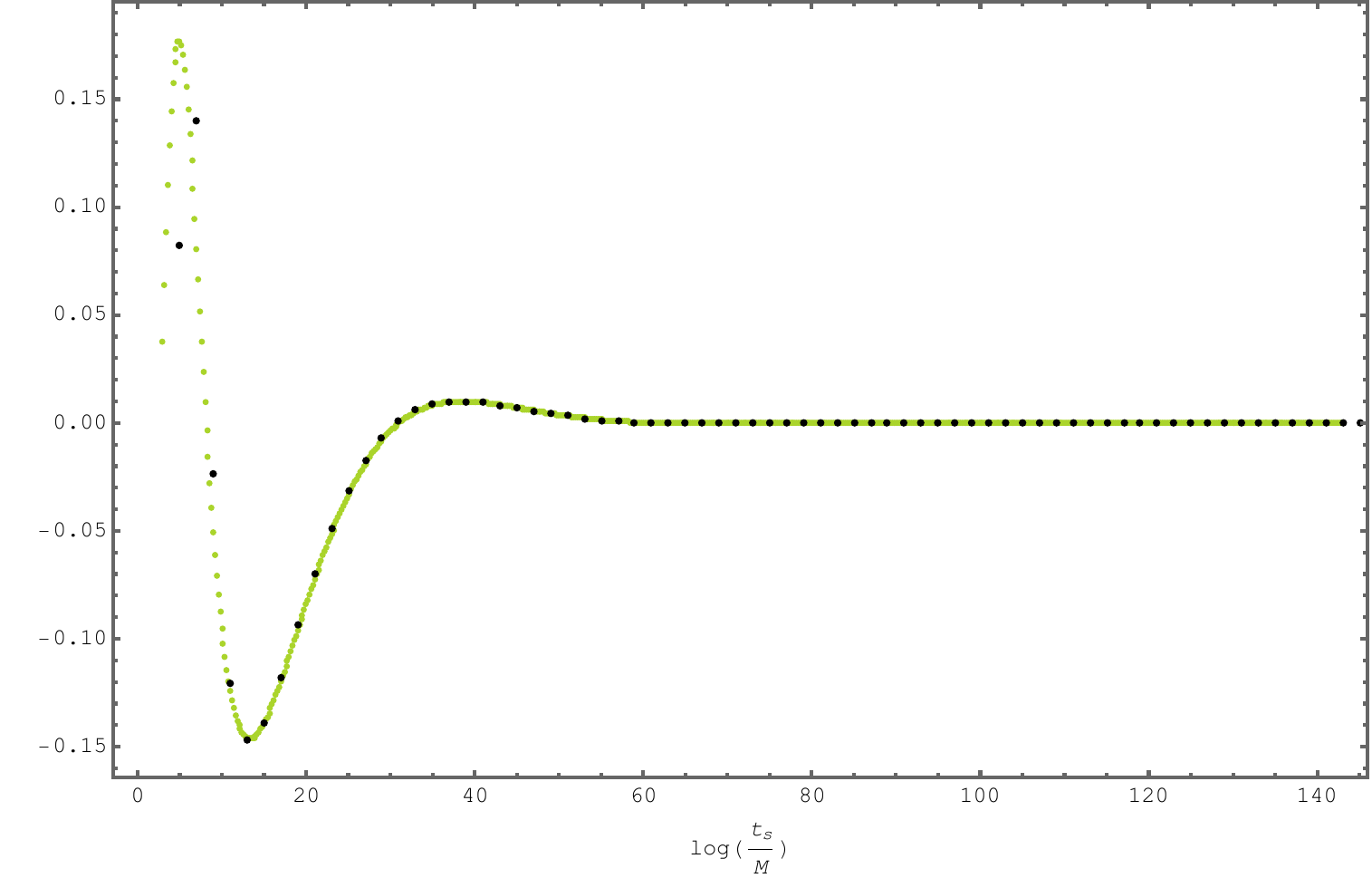}
\includegraphics[trim=0cm 0cm 0cm 0cm,clip=true,totalheight=0.22\textheight]{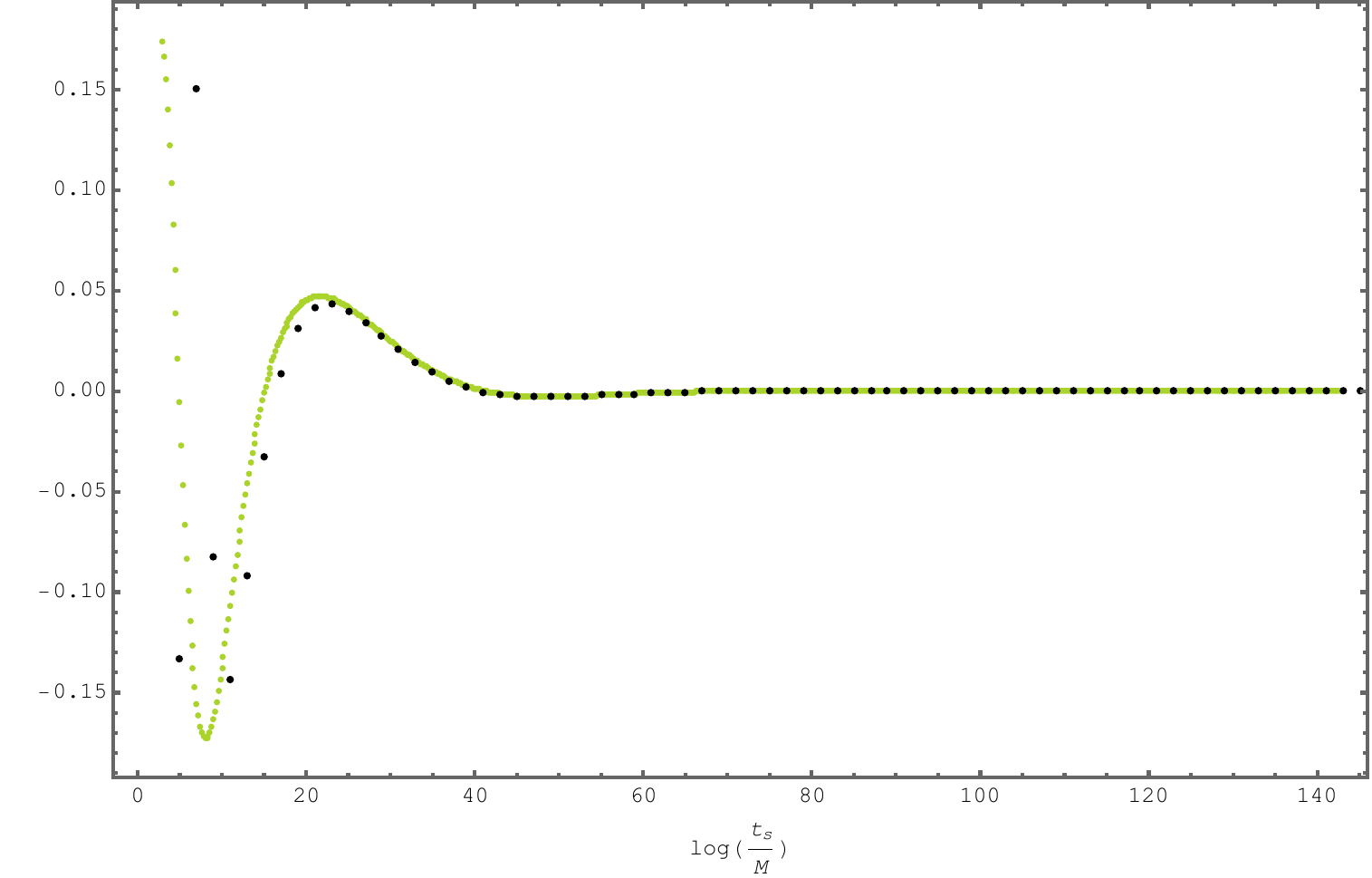}
\includegraphics[trim=0cm 0cm 0cm 0cm,clip=true,totalheight=0.22\textheight]{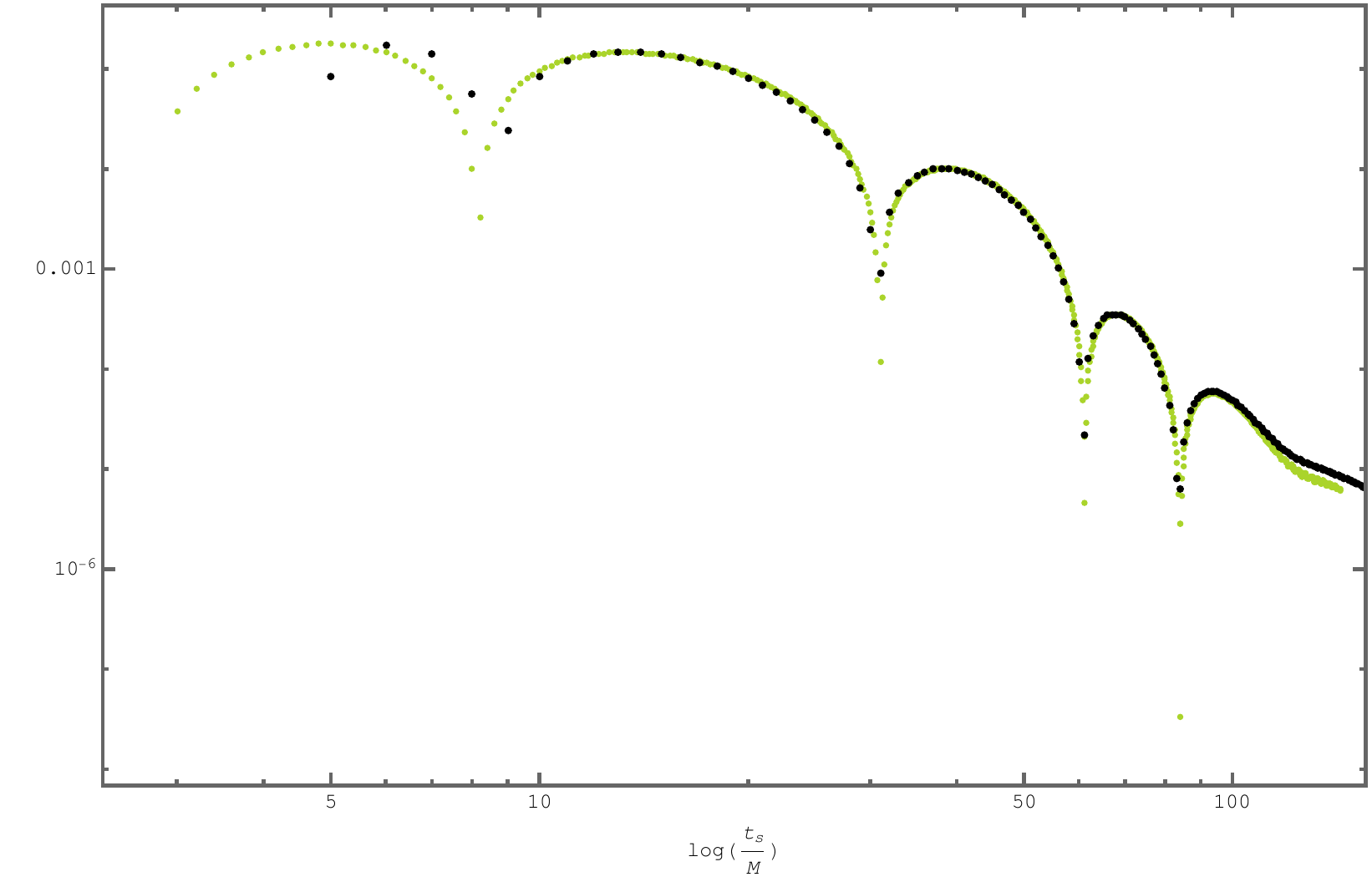}
\includegraphics[trim=0cm 0cm 0cm 0cm,clip=true,totalheight=0.22\textheight]{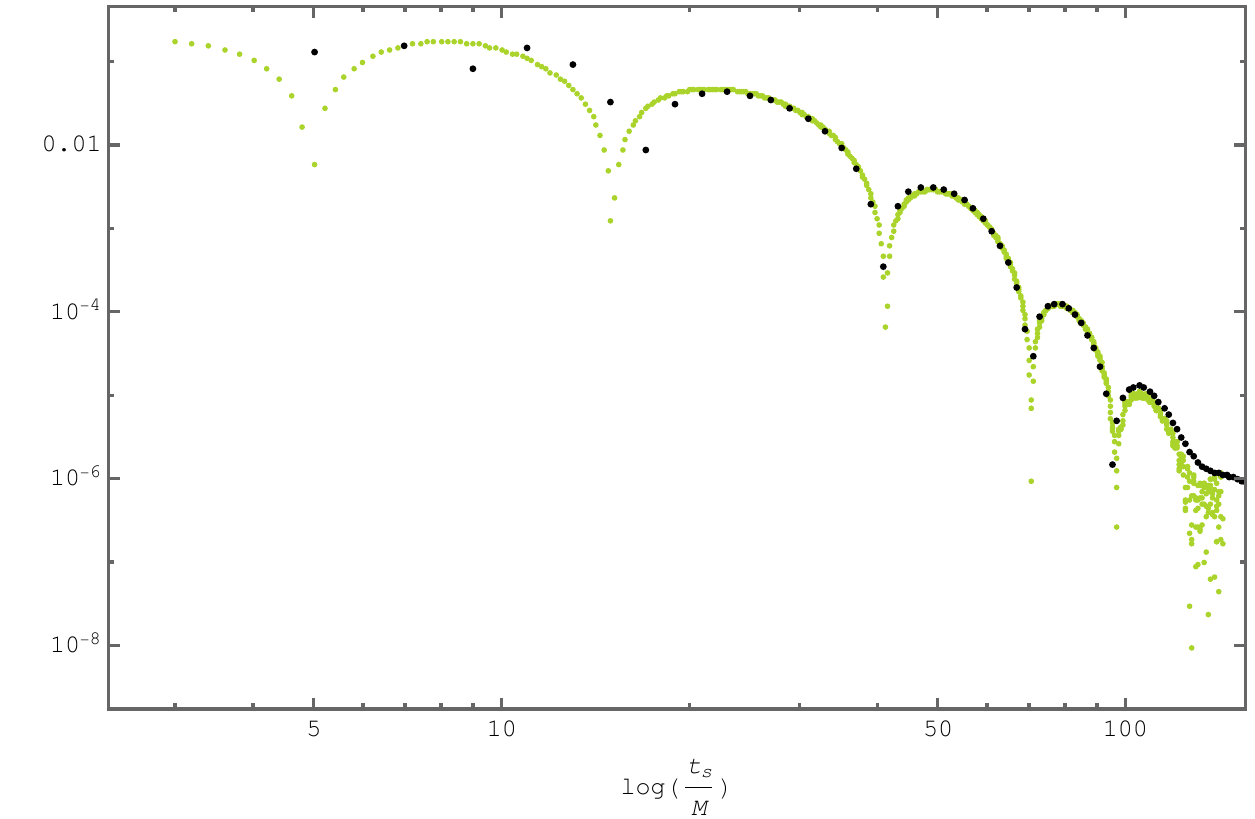}
\caption{The real (top-left) and imaginary (top-right) parts of $(\psi^{\text{in}}_\w)_u$  and $\psi^K_\w$  for the frequency $M\w=0.5$ and space point $r=3M$ are plotted as a function of the time $t_s$.  The corresponding log-log plots for the real parts (bottom-right) and imaginary parts (bottom-left) are also shown.}
\label{fig:I2}
\end{figure}
\section{Summary and Conclusions}
\label{sec:summary}

We have numerically computed spherically symmetric modes for a massless minimally coupled scalar field in both the {\it in} state for the collapsing null-shell spacetime and for the Unruh state in Schwarzschild spacetime.  The spherically symmetric {\it in} modes consist of sums of two different solutions to the mode equation that we call $f^{\rm in}_v$ and $f^{\rm in}_u$.  The Unruh state consists of the set of modes $f^{\mathscr{I}^-}$ that are positive frequency with respect to the usual time coordinate $t_s$ and originate on past null infinity along with the set of modes $f^K$ that are positive frequency with respect to the Kruskal time coordinate on the past black hole horizon.  With the exception of the $f^{\mathscr{I}^-}$ modes, each of these can be written as packets of certain modes that can be computed using separation of variables in the coordinates $t_s, \, r, \, \theta, \, \phi $.

The $f^{\mathscr{I}^-}$ modes can be computed using separation of variables and their time dependence is thus $e^{-i \w t}$.  Therefore, they oscillate at all times for fixed values of the spatial coordinates.  Inside the null shell the spherically symmetric {\it in} modes can be written as $f^{\rm in} = f^{\rm in}_v + f^{\rm in}_u$ with $f^{\rm in}_v \sim e^{-i \w v}$ and $f^{\rm in}_u \sim  - e^{-i \w u}$. These are separately solutions to the mode equation.  Outside the null shell, for fixed values of the spatial coordinates, the modes $f^{\rm in}_v$ approach the $f^{\mathscr{I}^-}$ modes at late times and the modes  $f^{\rm in}_u$ approach the $f^K$ modes at late times.  

Since the $ f^{\rm in}_v$ modes approach the $f^{\mathscr{I}^-}$ modes at late times, they also oscillate at all times.  However, the quantity $ f^{\rm in}_v - f^{\mathscr{I}^-}$ behaves very differently.  For fixed spatial points, both the real and imaginary parts undergo a few oscillations and then decrease like $t_s^{-3}$ with no apparent oscillations.  
This type of behavior also occurs at late times for the $f^K$ modes and the real part of the $ f^{\rm in}_u$ modes.  The oscillations occur for the imaginary part of the $ f^{\rm in}_u$ modes and the power law damping probably does as well.  However, our numerical calculations have not been done accurately enough to verify this.  

This quasinormal mode oscillations followed by a power law tail is exactly the type of behavior that has been found for perturbations due to a classical massless minimally coupled scalar field in Schwarzschild and other eternal black hole spacetimes~\cite{gcpp, burko-ori, barack}.  The connection with the quantum field theory calculations appears to be that the modes for $f^{\rm in}_v$, $f^{\rm in}_u$, and $f^K$ can be written in terms of packets of modes that are positive frequency with respect to $t_s$ and their complex conjugates.  The coefficients in these packets are determined by the behaviors of the $f^{\rm in}_v$, $f^{\rm in}_u$, and $f^K$ modes on the Cauchy surfaces where these coefficients are computed.  These behaviors are mathematically the same as initial data for noncompact perturbations.  

The fact that the approach of the {\it in} modes to the Unruh modes is via a power law in time at late times has significant implications for the behavior of the stress-energy tensor for the quantum field.  As discussed in Sec.~\ref{sec:numerical}, in the case of a collapsing null shell in 2D when the space coordinate is fixed, one finds that the $f^{\rm in}_u$ modes approach the $f^K$ modes exponentially in time.  The stress-energy tensor for the {\it in} state also approaches the one for the Unruh state exponentially in time~\cite{hiscock,Sandro-book}.  The difference between the two stress-energy tensors just consists of sums over modes; there are no renormalization counterterms involved.  Therefore it seems likely that, at least some components of the stress-energy tensor for the massless minimally coupled scalar field in the {\it in} state will have a power law approach to the corresponding components for the Unruh state. While the mechanism for the power law approach is not the same, this is similar to what was found for the radiation that a particle detector receives in the case of a black hole that forms from collapse~\cite{bh-evap-paper}.

\acknowledgments 

We would like to thank Emanuele Berti, Gregory Cook and Antoine Folacci for helpful conversations.  We would also like to thank Antoine Folacci for sharing some numerical data on quasinormal modes, Adam Levi for sharing some numerical data on solutions to the mode equation and helpful conversations about it, and Raymond Clark for providing power series solutions to the radial mode equation in Schwarzschild spacetime. This work was supported in part by the National Science Foundation under Grants  No. PHY-1505875, PHY-1912584 and PHY-2309186 to Wake Forest University. Some of the numerical work was done using the WFU DEAC Cluster; we thank the WFU Provost's Office and Information Systems Department for their generous support. A.F. acknowledges partial financial support by the Spanish Grant PID2023-149560NB-C21 funded by MCIN/AEI/10.13039/501100011033, and by the Severo Ochoa Excellence Grant CEX2023-001292-S.

    \end{document}